\documentclass[12pt,letterpaper,usenames,dvipsnames,svgnames]{article}
\usepackage{jheppub}

\usepackage{amscd} 
\usepackage{bm} 
\usepackage{dsfont} 
\usepackage{bbm} 
\usepackage{mathrsfs} 

\usepackage{tikz, pgf} 
\usetikzlibrary{arrows, calc, shapes, shapes.arrows, decorations.markings, decorations.pathmorphing}
\newcounter{sarrow}

     \tikzset{>=triangle 90}
     \tikzset{snake it/.style={-stealth, decoration={snake, 
        amplitude = .4mm, segment length = 2mm, post length=0.9mm},decorate}}
     \tikzstyle{gr}=[draw,circle,green!50!black,fill=green!50!black,scale=.6]

\usepackage{multirow} 
\usepackage{colortbl} 
\usepackage{arydshln} 

\usepackage{xcolor}

\def\nn{\nonumber}
\def\ph{\phantom}
\newcommand{\bpm}{\begin{pmatrix}}
\newcommand{\epm}{\end{pmatrix}}
\newcommand{\bsm}{\begin{smallmatrix}}
\newcommand{\esm}{\end{smallmatrix}}
\newcommand{\bspm}{\left(\begin{smallmatrix}}
\newcommand{\espm}{\end{smallmatrix}\right)}
\allowdisplaybreaks[2] 
\def\beq{\begin{equation}}
\def\eeq{\end{equation}}
\def\bar{\overline}
\def\til{\widetilde}
\def\hat{\widehat}

\def\v{\vee}
\def\^{\wedge}

\def\vev#1{{\langle{#1}\rangle}}

\def\Pf{{\rm Pf}}
\def\Im{{\rm Im}}

\def\im{{\rm im}}

\def\GL{{\rm GL}} 
\def\SL{{\rm SL}}

\def\SO{{\rm SO}}
\def\U{{\rm U}}
\def\SU{{\rm SU}}
\def\Sp{{\rm Sp}}

\def\so{\mathfrak{so}}
\def\u{\mathfrak{u}}
\def\su{\mathfrak{su}}
\def\sp{\mathfrak{sp}}

\def\C{{\mathbb C}} 
\def\cC{{\mathcal C}} 
\def\DD{{\mathbb D}} 

\def\he{{\hat e}}
\def\fg{{\mathfrak g}}
\def\tG{{\til G}}

\def\cL{{\mathcal L}}
\def\hm{{\hat m}}
\def\N{{\mathbb N}} 
\def\cN{{\mathcal N}}

\def\R{{\mathbb R}} 

\def\ft{{\mathfrak t}}
\def\hw{{\hat w}}
\def\Z{{\mathbb Z}}
\def\a{{\alpha}}

\def\b{{\beta}}
\def\g{{\gamma}}
\def\G{{\Gamma}}
\def\d{{\delta}}

\def\e{{\epsilon}}

\def\l{{\lambda}}
\def\L{{\Lambda}}
\def\m{{\mu}}
\def\n{{\nu}}

\def\r{{\rho}}
\def\s{{\sigma}}
\def\S{{\Sigma}}
\def\t{{\tau}}

\def\f{{\phi}}

\def\om{{\omega}}

\usepackage{tikz}
\usetikzlibrary{arrows,snakes,shapes.arrows,decorations.markings}
     \tikzset{>=triangle 90}
     \tikzstyle{bbc}=[draw,circle,fill=black,scale=.75]
     \tikzstyle{rc}=[circle,fill=red,scale=.6]
     \tikzstyle{wc}=[draw,circle,scale=.75]

\def\bar{\overline}
\def\til{\widetilde}
\def\hat{\widehat}
 
\def\vev#1{{\langle{#1}\rangle}} 

\def\v{\vee}
\def\^{\wedge}

\def\Im{{\rm Im}}

\def\U{{\rm U}}
\def\SU{{\rm SU}}
\def\SO{{\rm SO}}
\def\SL{{\rm SL}}
\def\GL{{\rm GL}} 
\def\Sp{{\rm Sp}}

\def\a{{\alpha}}

\def\b{{\beta}}
\def\g{{\gamma}}

\def\G{{\Gamma}}
\def\d{{\delta}}

\def\e{{\epsilon}}

\def\l{{\lambda}}

\def\L{{\Lambda}}

\def\m{{\mu}}
\def\n{{\nu}}

\def\r{{\rho}}
\def\s{{\sigma}}

\def\S{{\Sigma}}

\def\t{{\tau}}

\def\f{{\phi}}

\def\cC{{\mathcal C}}

\def\cL{{\mathcal L}}

\def\cN{{\mathcal N}}

\def\DD{\mathbb{D}} 

\def\N{\mathbb{N}}

\def\R{\mathbb{R}}

\def\beq{\begin{equation}}
\def\eeq{\end{equation}}
\def\nn{\nonumber}
\newcommand{\bpmat}{\begin{pmatrix}}
\newcommand{\epmat}{\end{pmatrix}}
\newcommand{\bsmat}{\begin{smallmatrix}}
\newcommand{\esmat}{\end{smallmatrix}}

\title{Dirac pairings, one-form symmetries and Seiberg-Witten geometries}
\author[a]{Philip C. Argyres,}
\author[b]{Mario Martone,}
\author[c]{Michael Ray}
\affiliation[a]{Physics Dept., U. Cincinnati, PO Box 210011, Cincinnati OH 45221, USA}
\affiliation[b]{Dept.\ of Mathematics, King’s College London, The Strand, London WC2R 2LS, UK}
\affiliation[c]{Dept.\ of Physics and Astronomy, Stony Brook U.,
Stony Brook, NY 11794-3840, USA}
\emailAdd{philip.argyres@gmail.com}
\emailAdd{mario.martone@kcl.ac.uk}
\emailAdd{michael.ray.1@stonybrook.edu}
\abstract{The Coulomb phase of a quantum field theory, when present, illuminates the analysis of its line operators and one-form symmetries. 
For 4d $\cN=2$ field theories the low energy physics of this phase is encoded in the special K\"ahler geometry of the moduli space of Coulomb vacua. 
We clarify how the information on the allowed line operator charges and one-form symmetries is encoded in the special K\"ahler structure.
We point out the important difference between the lattice of charged states and the homology lattice of the abelian variety fibered over the moduli space, which, when principally polarized, is naturally identified with a choice of the lattice of mutually local line operators. This observation illuminates how the distinct S-duality orbits of global forms of $\cN=4$ theories are encoded geometrically.}

\begin{document}
\maketitle


\section{Introduction}

A quantum field theory with a given spectrum and algebra of local operators may admit distinct ``global structures" encoded in a choice of extended (line, surface, etc.) ``probe" operators.  In the case of 4d Yang-Mills (YM) theories these global structures are identified with the global form of the gauge group together with additional discrete theta angles \cite{Aharony:2013hda, Gaiotto:2010be, Ang:2019txy}.  The choice of global structure is also related to the higher-form symmetries of the theory \cite{Gaiotto:2014kfa}.

In this paper we explain how this global structure is reflected in the geometry of the Coulomb branch of vacua of 4d $\cN{=}2$ field theories.
We will focus here on only the line operator spectrum and the 1-form symmetry of these theories.
A simple example to keep in mind is 4d $\cN{=}4$ supersymmetric YM (sYM) with gauge algebra $\su(2)$, which admits three compatible assignments of line-operator charges corresponding to three different global forms $\SU(2)$, $\SO(3)_+$ and $\SO(3)_-$.  
This theory has an exactly marginal coupling taking values in a ``conformal manifold".  
The three global structures form a single orbit under S-duality, meaning that upon traversing non-trivial loops in the conformal manifold, they are interchanged.  Finally, this theory has a $\Z_2$ 1-form symmetry under which the line operators may be charged.

This global structure data is completely encoded in the  \emph{charge lattice} and its \emph{Dirac pairing} which are physical attributes of the low energy theory in a Coulomb vacuum. One aim of this paper is to point out that then the global structure is encoded in the Coulomb branch geometry, and to explain this encoding.
This point has already been made in the context of lagrangian theories of class S, see \cite [sections 6.3 and 7.3.1]{Gaiotto:2010be}, \cite[section 5.4]{Tachikawa:2013hya}, and for $\SU(2)$ sYM  \cite[section 4.5]{Tachikawa:2013kta}.
We emphasize, however, that the notion of global structure interpreted in this way applies just as well to non-lagrangian theories where there is no gauge group.
We also point out that the encoding of the global structure in the Coulomb branch geometry cannot be as straightforward in general as it is in the class S examples where to each distinct global structure (i.e., ones not related by S-duality) corresponds a distinct Coulomb branch geometry. 
In particular, there are examples of theories whose different global structures are reflected in the same choice of CB geometries.


A characteristic feature of $\cN{=}2$ QFTs is that they have a Coulomb branch, an $r$-complex-dimensional manifold of $\u(1)^r$ gauge theory vacua with massive charged fields.  
The finite-energy states in these vacua carry electric and magnetic charges with respect to each of these $\u(1)$'s whose values form a rank-$2r$ lattice, the \emph{charge lattice} $\L$ of the theory.  
Choose some basis of this lattice with respect to which a charge $Q\in\L$ is represented by a $2r$-tuple of integers, $Q = ( p_1, \ldots, p_r , q_1 , \ldots , q_r)$.

The Dirac quantization condition \cite{Dirac:1931kp} implies that there is a non-degenerate integral pairing on the charge lattice, $J:\L\times\L\to\Z$, which can be written in a basis as $J(Q_1,Q_2) = Q_1 J Q_2^t$ where the right side is matrix multiplication with $Q_1$ and $Q_2$ interpreted as row vectors and $J$ as a $2r\times2r$ integer antisymmetric matrix.
If there is a basis of $\L$ in which $J$ has the $r\times r$ block form
\begin{align}\label{princ}
    J = \bpm & \ph{-}1 \\ -1 & \epm ,
\end{align}
i.e., a basis in which the Dirac pairings take their minimal allowed values, then the Dirac pairing is \emph{principal}.
But there is no reason that such a basis needs to exist, in which case the pairing is non-principal.
For example, most $\cN{=}4$ sYM theories have non-principal Dirac pairings.

The possible choices of global structure of the theory, and their associated 1-form symmetries, can be unambiguously read off from knowledge of the Dirac pairing on the charge lattice.
Indeed the different global structures correspond to the possible inequivalent refinements of the charge lattice on which the Dirac pairing extends to a principal pairing.
These refined lattices, $\cL$, are possible charges of a maximal mutually local set of line operators \cite{Gaiotto:2010be, Aharony:2013hda, Ang:2019txy}, so we refer to them as \emph{line lattices}.
This description of the global structures of an $\cN{=}2$ field theory does not depend on its having a gauge theory description, so applies equally well, for example, to isolated strongly-coupled $\cN{=}2$ superconformal field theories (SCFTs) for which there is no gauge group.

For $\cN{=}2$ SCFTs with exactly marginal couplings, the S-duality group acts on the global structures, organizing them into distinct orbits.
In the $\su(2)$ $\cN{=}4$ example mentioned above, all three global structures form a single S-duality orbit, but, by contrast, the 7 global structures of the $\su(4)$ $\cN{=}4$ theory form 2 separate S-duality orbits \cite{Aharony:2013hda}.

A feature of $\cN=2$ field theories, which was not particularly used in the original analysis of global structures \cite{Aharony:2013hda}, is that the low energy effective theory in the Coulomb vacua can be geometrically encoded in the special K\"ahler (SK) geometry of the theory's Coulomb branch \cite{Seiberg:1994aj, Seiberg:1994rs, Donagi:1995cf, Freed:1997dp}. 
The Seiberg-Witten curve --- a family of Riemann surfaces with a certain 1-form varying holomorphically over the Coulomb branch --- is a way of describing the SK structure.

The main goal of this paper is to clarify the extent to which the SK structure is sensitive to the global structures of the field theory. 
We will illustrate this in the relatively simple case of $\cN=4$ sYM theories, but our analysis applies generally to all $\cN=2$ theories. 
More specifically:
\begin{itemize}
    \item We clarify the distinction between two objects which are often conflated, the charge lattice $\L$ of the theory and the homology lattice $\L_X$ of the Seiberg-Witten curve.  In particular, if the intersection pairing on $\L_X$ is principal, then $\L_X$ is naturally identified with a choice of line lattice $\cL$ rather than with the charge lattice $\L$.
    
    \item We show that a given Coulomb branch K\"ahler geometry can admit a discrete set of distinct SK ``models'' which differ by the symplectic pairing on their homology lattices.
    
    \item We point out that the S-duality group which is discernible from the Coulomb branch geometry depends on the SK model.  In particular, the S-duality group ``visible" in the SK model in which the homology lattice is the charge lattice is larger than the physical S-duality group as it does not distinguish between the different global structures of the theory.  Thus, for example, these ``coarse" S-duality groups are $\SL(2,\Z)$ for all the $\cN{=}4$ sYM theories.
\end{itemize}
The second point, in particular, has been made before in the context of class-S examples in  \cite{Gaiotto:2010be, Tachikawa:2013hya, Tachikawa:2013kta}, as mentioned above.
This point would seem to predict that (a) there are different principally-polarized ``SK models'' of a Coulomb branch corresponding to the different  global structure S-duality orbits; (b) there is a unique model with maximally non-principal pairing whose homology lattice equals the charge lattice; and (c) there will exist models corresponding to non-maximal choices of line lattices with pairings whose invariant factors divide those of the maximally non-principal pairing model.
In fact, however, it seems that prediction (a) is incorrect, as we will point out a counter-example shortly.

As part of the exposition of the above points, we re-derive some results which, while known to experts, may not be more widely appreciated.  In particular, we derive from well-known semiclassical field theory facts the (non-principal) Dirac pairing on the charge lattice of $\cN=4$ sYMs. These results are consistent with those of \cite{DelZotto:2022ras}.  
Also, we show how to re-derive the results of \cite{Aharony:2013hda} by counting maximal symplectic sublattices, and present simple algorithms for doing so.

In this work we only apply our analysis to re-derive the global structures and one-form symmetries of $\cN=4$ sYM theories as examples, but our techniques apply more generally to $\cN=2$ theories for which many results are by now known \cite{Closset:2020scj, DelZotto:2020esg, Closset:2020afy, Bhardwaj:2021pfz, Hosseini:2021ged, Buican:2021xhs, Closset:2021lwy, Bhardwaj:2021mzl, Carta:2022spy, DelZotto:2022ras}. 
In particular, an immediate implication of the known classification of rank-1 Coulomb branch geometries \cite{Argyres:2015ffa, Argyres:2015gha, Argyres:2016xua, Argyres:2016xmc} is that since --- with one exception --- all have only principally polarized homology lattices, it follows that all rank-1 SCFTs have principal Dirac pairings, their charge and line lattices coincide, and they have no 1-form symmetries.
This agrees with the predictions from string constructions and BPS quivers \cite{Closset:2020scj, DelZotto:2020esg, Closset:2020afy, Bhardwaj:2021pfz, Hosseini:2021ged, Buican:2021xhs, Closset:2021lwy, Bhardwaj:2021mzl, Carta:2022spy, DelZotto:2022ras}.
The one exception is the Coulomb branch geometry of the $\cN{=}2^*$ $\su(2)$ sYM theory which also has a model with non-principally polarized homology lattice, which is consistent with the field theory expectation.
We describe this example in detail in section \ref{su2curves} below.

\paragraph{$\cN{=}3$ Coulomb branches which to not admit principal polarizations.}

A class of theories for which the derivations presented here could be even more directly applied are $\cN{=}3$ theories \cite{Garcia-Etxebarria:2015wns, Aharony:2015oyb, Aharony:2016kai}.
In fact the moduli spaces of $\cN{=}3$ theories closely resemble those of the $\cN{=}4$ theories described here, since both, as complex geometries, are complex orbifolds by reflection groups.
For $\cN{=}4$ theories the orbifold groups are the Weyl groups of the gauge algebras (which are crystallographic real reflection groups), which are substituted by the more general crystallographic complex reflection groups in $\cN{=}3$ theories \cite{Caorsi:2018zsq,  Bonetti:2018fqz,  Argyres:2019ngz}. 
This suggests that a perhaps straightforward generalization of the analysis in this paper will allow the computation of one-form symmetries of all $\cN{=}3$ theories in a uniform way.

Not all crystallographic complex reflection group orbifolds admit SK models for which the homology lattice is principally polarized.
For example, at rank 2 the exceptional complex reflection group $G_8$ in the Shephard-Todd classification \cite{Shephard:1954} (with Coulomb branch scaling dimensions 8 and 12) only admits integer symplectic forms with minimum invariant factors (1,2), so are never principal; see section 3.2 of \cite{Caorsi:2018zsq}.
Furthermore, this Coulomb branch arises as the moduli space of an $\cN{=}3$ SCFT with a known M-theory construction \cite{Kaidi:2022lyo, Garcia-Etxebarria:2016erx}.
The counting of global structures of theories with non-principal Dirac pairing, reviewed below, implies that this theory has at least 3 distinct global structures.
But the absence of an SK model with principal polarization means that no Coulomb branch geometries exist which realize a choice of maximal line lattice for this theory.
It remains to be understood what the implications of this fact are and whether the absence of a principally polarized model for the Coulomb branch of this theory has an interpretation in terms of its spectrum of extended operators or its generalized symmetry. This questions certainly deserves further study.\\

The rest of the paper is organized as follows.  Section \ref{sec2} reviews the definitions of and relations between the charge lattice, the Dirac pairing, and the possible line lattices.  We also review how these are related to the 1-form symmetry.  Section \ref{sec3} begins the discussion of $\cN=4$ sYM theories by deriving their charge lattices and Dirac pairings. Finally, section \ref{sec4} discusses how the global structure (or, choice of line lattice) appears in the SK structure of the Coulomb branch.  We conclude the section with an explicit discussion of low-rank examples.  There are also three technical appendices summarizing some basic properties of simple Lie algebras, discussing how to bring symplectic pairings to canonical form, and showing how to count inequivalent symplectic sublattices.

\section{Charge lattices in 4d QFTs}\label{sec2}

Associated to each quantum field theory with a Coulomb phase, there is a linear space of allowed possible deconfined gauge charges of finite energy states.  
In 4d these are electric and magnetic charges.  
A vacuum is a Coulomb phase if there is a low energy $\U(1)^r$ gauge theory with mass gap for all charged states; $r$ is the \emph{rank} of the vacuum.
Furthermore, we assume that there are states carrying electric and magnetic charges with respect to each $\U(1)$ factor, so the linear space of possible charges has real dimension $2r$.%
\footnote{We exclude IR free non-abelian vacua or ones with non-compact $\U(1)$ gauge factors from our definition of Coulomb phase.}

Then the set of charges of finite-energy states which occur in the theory form a rank-$2r$ {\bf charge lattice}, $\L \simeq \Z^{2r}$. 
These include the charges of single-particle states, though there is no requirement that all lattice charges correspond to single-particle states.
The physically occurring charges form a lattice because if localized states%
\footnote{By localized in this context we mean only that the energy density falls off at least as fast as $\r(t,\vec x) \sim |\vec x|^{-4}$ as $|\vec x|\to\infty$.}
with charges $p$ and $q$ occur, then there is a localized finite energy state with charge $p+q$ (e.g., an approximate 2-particle state with the particles sufficiently far apart).
Furthermore, all charges are commensurate by the Dirac quantization condition \cite{Dirac:1931kp, Schwinger:1966nj, Zwanziger:1968ams, Zwanziger:1968rs}, which implies there is a \emph{Dirac pairing}, a non-degenerate, antisymmetric, bilinear map $J:\L \times \L \to\Z$.
It supplies the charge lattice with a symplectic structure.

Note that the normalization of $J$ is not a matter of convention.
For instance, the value of the Dirac pairing between a pair of (sufficiently localized) dyonic states measures the angular momentum carried by their electromagnetic field \cite{Coleman:1982cx} in units of $\hbar/2$.
With a common definition of electric and magnetic charges --- $e = \int_{S^2} *F$ and $g = (4\pi)^{-1} \int_{S^2} F$ --- the Dirac quantization condition reads $eg=n/2$, $n\in\Z$.  
We are implicitly using a different normalization of charges --- e.g., by defining magnetic charge to be $g = (2\pi)^{-1} \int_{S^2} F$ --- in which $J$ is integral.  This  does not change the fact that the normalization of $J$ has a definite physical meaning.

With respect to a particular basis of the lattice, $\{ e_1, \ldots, e_{2r} \}$, the Dirac pairing is represented by a non-degenerate antisymmetric $2r\times 2r$ matrix $J_{ab} = J(e_a, e_b) \in\Z$.
There exists a \emph{symplectic basis} in which $J$ is skew-diagonal, i.e., is of the form
\begin{align}\label{sympbas}
    J &= \bpm & \ph{-}D \\ -D & \epm & 
    &\text{with}& 
    D &= {\rm diag} \{d_1, \ldots, d_n\} , &
    d_i &\in \N.
\end{align}
Furthermore, one can choose this basis so that 
\begin{align}\label{ddiv}
    d_i \mid d_{i+1} .
\end{align}
In this case the $d_i$ are the unique \emph{invariant factors} of the Dirac pairing, though the symplectic basis is not unique.
If $d_i=1$ for all $i$ we say the Dirac pairing is \emph{principal}.
The existence and uniqueness of this invariant factor decomposition is ensured by the structure theorem for finitely generated modules over a principal ideal domain, and is reviewed --- together with an algorithm for computing the invariant factors --- in appendix \ref{invfacapp}.

For any electric and magnetic charges, whether they are in the charge lattice or not, a ``probe'' Wilson-'t Hooft line operator can be defined by specifying appropriate boundary conditions on the behavior of the electric and magnetic fields as one approaches the line \cite{Kapustin:2005py}.
The electric and magnetic charges of Wilson-'t Hooft line operators are $\u(1)_E \oplus \u(1)_M$ 1-form symmetry charges for each IR gauge $\u(1)$ factor \cite{Gaiotto:2014kfa}.
The set of these line operator charges should include charges in the charge lattice, $\L$, since in the IR limit any massive charged state can be approximated by the insertion of a line operator located at the world-line of its center of mass \cite{Wilson:1974sk}.
The charges of probe lines are restricted to lie in a lattice by Dirac quantization, which, for line operators, is the condition that they are mutually local or ``genuine" line operators \cite{Kapustin:2014gua}, i.e., that they are not the boundary of a topological surface operator that can be detected by other genuine line operators.  
(For example, these topological surface operators are the world-volume of the Dirac strings emanating from the world-line of a magnetic monopole.)
Thus the charges of genuine probe lines must have integral Dirac pairing with charges in $\L$ \cite{Gaiotto:2010be}.

This suggests defining the Dirac-dual lattice, $\L^J$, with the interpretation as the lattice of ``possible'' probe Wilson-'t Hooft line operator 1-form charges.  
It is the maximal lattice in $\R\otimes_\Z \L$ such that \begin{align}\label{dualL}
    J(\L^J,\L) \in \Z.
\end{align}  
In other words it is integrally ``dual" to $\L$ with respect to the Dirac pairing.
The charge lattice is a sublattice of the dual lattice of index 
\beq
|\L^J/\L| = (\Pf J)^2\ {\rm with}\ \Pf J = \det D = \prod_i d_i, 
\eeq
so $\L$ is a proper sublattice of $\L^J$ if and only if $J$ is non-principal.
In terms of the symplectic basis $\{ e_a, \ a=1, \ldots, 2r \}$ in \eqref{sympbas}, the symplectic basis basis of $\L^J$ is $\{ d_i^{-1} e_i, d_i^{-1} e_{i{+}r}, \ i=1, \ldots, r \}$

If $\L$ is a proper sublattice of $\L^J$ then the extension of the Dirac pairing to $\L^J$ will not be integral, i.e., $J(\L^J,\L^J) \notin \Z$.
Indeed, in the symplectic basis given above, 
\begin{align}
    J = \bpm &\ph-D^{-1} \\ -D^{-1} & \epm
    \qquad \text{on \ \  $\L^J$.}
\end{align}
Define a {\bf line lattice}, $\cL \subset \L^J$, to be a maximal sublattice of $\L^J$ such that $J(\cL,\cL) \in \Z$ \cite{Gaiotto:2010be}.
A line lattice is thus interpreted as a maximal lattice of mutually local Wilson-'t Hooft probe line operators of the low energy $\U(1)^r$ gauge theory.
It follows from its definition that the extension of the Dirac pairing to $\cL$ is principal, and that 
\beq
|\L^J  / \cL| = |\cL / \L| = \Pf J.
\eeq
There can be finitely many inequivalent line sublattices; the number depends in a complicated way on the prime decomposition of the invariant factors, $d_i$, of the Dirac pairing; some simple examples are computed in appendix \ref{appC}.

In the case of gauge theories, the choice of line lattice is associated to the choice of global structure (including ``discrete theta angles'') of the theory \cite{Aharony:2013hda}.
The connection between the choice of line lattices and this global structure arises as follows.
In gauge theories for which there exist local operators in a Coulomb vacuum which create states carrying (gauge) charges in $\L$, Wilson-'t Hooft line operators carrying charges in $\L$ can be screened.
That is, there are gauge-invariant line segment operators consisting of Wilson-'t Hooft lines which end on these local operators.
In that case the $(U(1)_E \times U(1)_M)^r$ 1-form group symmetry is broken to a discrete subgroup consisting of those group elements which act trivially on all lines in $\L$, since the $S^2$'s which carry the 1-form topological symmetry operators no longer link the line segments.
If the minimum charge in $\L$ with respect to a given $U(1)$ 1-form symmetry factor is $d$, then that factor is broken (at least) to its $\Z_d$ subgroup.
Therefore, in a basis corresponding to the invariant factor decomposition \eqref{sympbas}, the 1-form symmetry group is broken (at least) to the finite subgroup, known as the 1-form defect group \cite{Tachikawa:2013hya,DelZotto:2015isa,Albertini:2020mdx}: $\DD^{(1)} = \bigoplus_{i=1}^r \left( (\Z_{d_i})_E \oplus (\Z_{d_i})_M \right)$, where, again, the $d_i$s are the invariant factors in \eqref{sympbas}. But, like the Dirac-dual lattice $\L^J$, the charges allowed in this group generally do not have integral Dirac pairing, so the actual 1-form group is a maximal subgroup of $\DD^{(1)}$ mutually local charges (which has cardinality $\prod_i d_i$).  
The choice of this subgroup is the global structure computed in \cite{Aharony:2013hda,DelZotto:2022ras}.
The above description in terms of maximal Dirac-local subgroups of the finite group $\DD^{(1)}$ can be recast in terms of $\cL \subset \L^J$ maximal Dirac-local sublattices.
Since $\L \subset \cL \subset \L^J$, we have the associated finite group inclusions $\cL/\L \subset \L^J/\L$, and, by working in the symplectic basis, it is easy to see that  $\L^J/\L = \DD^{(1)}$.

An immediate consequence is that if there is a non-trivial global structure or 1-form symmetry, then the Dirac pairing on the charge lattice is not principal.  So from this perspective, and the results of \cite{Aharony:2013hda}, it is now obvious that $\cN=4$ sYM theories will generally have non-principal Dirac pairings. In the next section we show how to compute the Dirac pairings directly in the field theory.

The results derived in the next section agree with those which appeared in \cite{DelZotto:2022ras}. 
In the derivation of \cite{DelZotto:2022ras}, though, the Dirac pairing appears in a more indirect way, via the \emph{symmetry TFT} \cite{Apruzzi:2021nmk,Apruzzi:2022dlm}, i.e., the topological sector of the non-invertible field theory whose boundary theory is the four dimensional theory.

\section{Non-principal Dirac pairings for $\cN{=}4$ super Yang-Mills} \label{sec3}

In the case of an $\cN{=}4$ sYM theory with simple gauge algebra $\fg$, in a Coulomb vacuum the electric charges span the root lattice $\G_r$ of $\fg$ since at weak coupling all fields are in the adjoint representation. 
A semiclassical analysis shows that magnetic monopole charges span the co-root lattice $\G_r^\v$, which is the ``magnetic" root lattices of of $\fg^\v$, the GNO or Langlands dual of $\fg$ \cite{Goddard:1976qe, Witten:1978mh, Osborn:1979tq, Kapustin:2005py}.  
(See appendix \ref{Cartmat} for a review of simple Lie algebra definitions.)
So the charge lattice is
\begin{align}\label{chrglatt}
    \L = \G^\v_r \oplus \G_r ,
\end{align}
the span of the electric and magnetic sublattices.
These sublattices are each lagrangian with respect to $J$, so the Dirac pairing is determined by the $r\times r$ pairing $B := J(\G_r,\G^\v_r)$ between the two.  
And any integral non-degenerate $B$ defines a potential Dirac pairing on $\L$.

The Dirac pairing is fixed up to normalization by demanding it be Weyl invariant.  
Recall that the Weyl group is a discrete subgroup of the gauge group which acts on the charge lattice, so any physical observable which is constant on the Coulomb branch and so independent of the adjoint Higgs vev --- such as the value of the Dirac pairing between a pair of (sufficiently localized) charged states --- must be invariant under the Weyl group.

The Weyl group, $W$, of $\fg$ and $\fg^\v$ are the same.  
Given its linear action on $\G_r$, the magnetic Weyl action is determined by the semiclassical description of magnetic monopoles \cite{Goddard:1976qe, Kapustin:2005py}.
Indeed, by GNO duality, the magnetic root lattice is the dual of the ``electric" weight lattice, $\G^\v_r \simeq (\G_w)^*$,%
\footnote{Here the dual $(\G_w)^*$ means the space of linear maps of $\G_w$ to $\Z$, and should not be confused with the notion of ``dual" with respect to the Dirac pairing used in \eqref{dualL}.}
and so inherits a dual action of the Weyl group.%
\footnote{Some Weyl groups have non-trivial outer automorphisms so one might worry that there are inequivalent choices of how the electric and magnetic Weyl actions are put into correspondence.
Weyl groups are comprised of rotations and reflections with respect to the Killing metric on the weight space, which can be characterized as the elements having positive and negative determinants, respectively, when realized as matrices in some basis.  
Thus the dual of reflections are also reflections. 
All automorphisms of Weyl groups which preserve the set of reflections are either inner or are Dynkin diagram automorphisms \cite{Franzsen:2001}.
In either case they are equivalent to a permutation of a basis of simple roots, and so are equivalent up to a choice of lattice basis.}
Let $\ft\subset\fg$ be a Cartan subalgebra and $\ft^*\subset\fg^\v$ the dual Cartan subalgebra (or weight space); so $\G_r \subset \ft^*$ and $\G^\v_r \subset \ft$.
Then the $W$ action on these lattices is generated by reflections $\s_k$ and $\s_k^\v$ associated to each simple root $\a_k$, $k=1,\ldots,r$, whose actions are given in \eqref{Weylact}, so $\s_k\in W$ acts on $\L = \G^\v_r \oplus \G_r$ as $\s^\v_k \oplus \s_k$.

Invariance of the Dirac pairing is the condition that
\begin{align}\label{Cartan1}
    B_{ij} := J(\a_i,\a^\v_j) &= J(\s_k(\a_i),\s^\v_k(\a^\v_j)) \quad \text{for all} \quad k = 1, \ldots, r .
\end{align}
Using the action \eqref{Weylact} then implies that
\begin{align}\label{Cartan2}
    \a_i(\a^\v_k) \, B_{kj} + B_{ik} \, \a_k(\a^\v_j)
    = \a_i(\a^\v_k) \, B_{kk} \, \a_k(\a^\v_j)
\end{align}
for all $i$, $j$, $k$.  Choosing $i=j$ and $i=k$ then implies that $B_{ii}=2n$ for some $n$ and for all $i$, and that $B_{ij} = n \ \a_i(\a^\v_j)$.
The Cartan matrix of $\fg$, $A_\fg$, is defined to have matrix elements $(A_\fg)_{ij} = \a_i(\a^\v_j)$.
The Cartan matrices of simple Lie algebras are recalled in appendix \ref{Cartmat}.

Thus invariance under the Weyl action determines the Dirac pairing up to an overall normalization.
And we have found that in a particular basis (corresponding to simple co-roots and roots) the Dirac pairing of the $\cN{=}4$ sYM with (electric) gauge algebra $\fg$ is
\begin{align}\label{Jfg}
    J_\fg &= 
    n\, \bpm 0 & (A_\fg)^t \\ - A_\fg & 0 \epm  = n\, \bpm 0 & A_{\fg^\v} \\ - A_\fg & 0 \epm, &
    n &\in \N ,
\end{align}
since $A_{\fg^\v} = (A_\fg)^t$, and $n$ is a non-zero integer because of the integrality of the Dirac pairing and the fact that the entries of the Cartan matrix have no common divisor.  (The sign of $J_\fg$ is conventional.)

The normalization can be determined by constructing electrically and magnetically charged states at weak coupling (i.e., semiclassically) with minimum non-zero Dirac pairing.
In the case of Yang-Mills theory with adjoint matter the Dirac pairing between the W-boson and 't Hooft-Polyakov monopole corresponding to an $\su(2)$ subalgebra associated to a root is 2 in the units used here \cite{tHooft:1974kcl, Polyakov:1974ek, Coleman:1982cx}.  
This implies that $n=1$ in \eqref{Jfg}, since diagonal elements of $A_\fg$ are 2.

This result also agrees with that found from the  BPS quiver description of the BPS spectrum as reported in \cite{DelZotto:2022ras}.%
\footnote{The Dirac pairing shown there has the form $J = \bspm A-A^t & A^t\\ -A & 0\espm$ which is related to \eqref{Jfg} by a change of lattice basis by the $r\times r$ block $\GL(2r,\Z)$ matrix $\bspm 1&0\\1&1\espm$.}

With the Dirac pairing in hand, we now determine its invariant factors --- the $d_i$ in \eqref{sympbas} and \eqref{ddiv} --- that invariantly characterize it.  
One can determine them by direct computation, as outlined in appendix \ref{invfacapp}.
They are simply the diagonal elements of the Smith normal form of the Cartan matrix, which are
\begin{align}\label{N4invfac}
\begin{array}{ll|cccccc}
\fg & & d_1 & d_2 & \cdots & d_{r-2} & d_{r-1} & d_{r} \\
\hline
\su(r{+}1) & r\ge1 & 1 & 1 & \cdots & 1 & 1 & r{+}1 \\        
\so(2r{+}1)\ & r\ge1 & 1 & 1 & \cdots & 1 & 1 & 2 \\        
\sp(2r) & r\ge1 & 1 & 1 & \cdots & 1 & 1 & 2 \\        
\so(2r) & r\ge3 \ \text{odd} & 1 & 1 & \cdots & 1 & 1 & 4 \\        
\so(2r) & r\ge2 \ \text{even} & 1 & 1 & \cdots & 1 & 2 & 2 \\        
E_r & r=6,7,8 & 1 & 1 & \cdots & 1 & 1 & 9{-}r \\        
F_4 & & 1 & 1 & \cdots & 1 & 1 & 1 \\        
G_2 & & 1 & 1 & \cdots & 1 & 1 & 1     
\end{array}
\end{align}
Note that the only Dirac pairings which are principal are those of the $E_8$, $F_4$, and $G_2$ gauge theories.
By the structure theorem (see appendix \ref{invfacapp}) which ensures the existence and uniqueness of the canonical symplectic structure \eqref{sympbas} with invariant factors satisfying the divisibility conditions \eqref{ddiv}, and by the fact that from \eqref{Jfg} $\det A_\fg = \prod_i d_i$, the determinants of the Cartan matrices, listed in appendix \ref{Cartmat}, completely determine the invariant factors listed in \eqref{N4invfac} without further computation except in the cases of $\fg=\su(r{+}1)$ for those $r{+}1$ not square-free and of $\so(2r)$.  
We determined these last two cases by direct computation for $r < 20$, but have not worked out a proof for all $r$.

As noted in the last section, the number of inequivalent line lattices following from these Dirac pairings should equal the number of different global structures for these Yang-Mills theories as computed in \cite{Aharony:2013hda}.
This equality follows from the observation that $\oplus_{i=1}^r \Z_{d_i}$ in \eqref{N4invfac} is the center of the simply connected Lie group of $\fg$, and from the connection between the counting of maximal Dirac-local symplectic sublattices of the Dirac-dual lattice and the counting of maximal Dirac-local subgroups of the defect 1-form group.
In appendix \ref{appC} we do the sublattice counting directly, verifying the equality.

\section{Comparison to CB geometry constructions}\label{sec4}

We now discuss how non-principal Dirac pairing on the charge lattice and the choice of line lattice appear in the Coulomb branch geometry of $\cN{=}2$ field theories.  

The Coulomb branch $\cC$ --- the moduli space of Coulomb vacua --- is a special K\"ahler (SK) space \cite{Donagi:1995cf, Freed:1997dp}.  
The SK structure on a rank-$r$ Coulomb branch can be expressed in terms of a family of rank-$r$ abelian varieties $X_u$, $u\in\cC$, varying holomorphically over $\cC$.
$X_u$ often appears as a sub-variety of the Jacobian variety of the Seiberg-Witten curve $\S_u$ in terms of which some SK geometries are written; see, e.g.,  \cite{Donagi:1995cf, Donagi:1997sr} and references therein.
The abelian varieties carry a choice of polarization, which can be thought of as a choice of integral symplectic pairing $P:\L_X\times\L_X\to\Z$ on their lattice of homology 1-cycles $\L_X := H_1(X_u,\Z) \simeq \Z^{2r}$.
$P$ is the pairing induced by the intersection pairing on homology 1-cycles on the Seiberg-Witten curve.
We will refer to $\L_X$ as the {\bf homology lattice} and $P$ as its {\bf polarization} which should not be confused with the charge and line lattice and their Dirac pairings. In fact our focus will be on understanding the relationship among these objects.

The main conclusion is that the homology lattice and polarization appearing in the SK structure of the Coulomb branch need not coincide with the charge lattice and Dirac pairing of the field theory.
Indeed, there is a discrete set of closely-related SK structures compatible with a given Coulomb branch K\"ahler geometry, in which (homology lattice, polarization) can take values ranging from the (charge lattice, Dirac pairing) to various (line lattices, principal pairings) as well as intermediate possibilities.
This choice of SK structure can thus be used to encode a choice of global structure of the field theory.

We illustrate this in the context of $\cN{=4}$ sYM theories. Specifically, we examine some examples of Coulomb branch geometries of $\cN{=}2^*$ theories appearing in the literature and show how to  determine how their homology lattices and polarizations are related to the field theory charge lattices and Dirac pairings.
These cases have an exactly marginal coupling and so S-duality groups related to the topology of their conformal manifolds.
We discuss how the S-duality group visible from the Coulomb branch geometry is related to that of the field theory.

\subsection{Review of $\cN=4$ moduli space geometry}\label{sec41}

The moduli space of $\cN{=}4$ sYM with gauge algebra $\fg$ is the flat orbifold geometry \cite{Seiberg:1997ax}:
\beq
\mathcal{M}_\fg\equiv\C^{3r}/W_\fg.
\eeq
Here $\s_k\in W_\fg$ acts as $\s_k: \vec z\otimes\m \mapsto \vec z\otimes \s_k(\m)$ where $\vec z\otimes\m \in \C^3 \otimes_\R \ft^* \simeq \C^{3r}$ with $\ft\subset \fg$ is a Cartan subalgebra and the $\s_k$ action on $\ft^*$ given by \eqref{Weylact}.
Choosing an $\cN{=}2 \subset \cN{=}4$ subalgebra, the associated Coulomb branch ``slice", $\cC_\fg$, is the orbifold $\cC_\fg \doteq \C^r/W_\fg$ where $\s\in W_\fg$ acts on $\C^r \simeq \C \otimes_\R \ft^*$ as above. 
The complex scaling action of the spontaneously broken scale plus R-symmetries is diagonalized by a (any) algebraic basis of Weyl-invariant polynomials in the complex coordinates on $\C^r$, $\{ u_1, \ldots, u_r\}$, whose degrees are the exponents plus one of the Weyl group.%
\footnote{The exponents of Weyl group $W_\fg$ are the $r$ integers $0<e_i<h$ with $\gcd(e_i,h)=1$ where $h$ is the Coxeter number of $\fg$ \cite{Humphreys:1990}.}  
Since Weyl groups act as (real) complex reflection groups on $\C^r$, the Chevalley-Shepherd-Todd theorem \cite{Shephard:1954, Chevalley:1955} implies that $\cC_\fg \simeq \C^r \ni (u_1,\ldots,u_r)$ as a complex space.

This metric and complex structure by themselves do not specify the SK geometry on $\cC_\fg$.
An SK structure on $\cC_\fg$ can be specified by choosing a holomorphic section, $s$ --- the ``special section" --- of a flat rank $2r$ complex vector bundle over $\cC_\fg$ with structure group $\Sp_P(2r,\Z)$ \cite{Freed:1997dp, Argyres:2019ngz, Argyres:2019yyb}.  
Its fibers are the complexification of the (linearly) dual homology lattice, $\C \otimes_\Z \L_X^* \simeq \C^{2r}$, so inherit a constant symplectic form from the polarization $P$ of $\L_X$.
$s$ satisfies the integrability condition $J(ds \, \overset{\^}{,} \, ds)=0$ where $d$ is the exterior derivative on $\cC_\fg$, and the (positive) metric on $\cC_\fg$ is given by $ds^2 = i J(ds , d\bar s)$.
In the $\cN{=}4$ case the components of $s$ are locally flat coordinates on $\cC_\fg$ which vanish at the origin (the conformal vacuum) \cite{Argyres:2019ngz, Argyres:2019yyb}.
The physical significance of the special section is that the dual pairing $\L_X^* \times \L_X \to \Z$ induces the central charge map $Z: \cC_\fg \times \L_X \to \C$ where $Z(u,q) := s(u)(q)$ and whose norm is the BPS mass of a state of EM charge $q$ in the vacuum $u$.

Choose a basis $(m_i, e^i)$, $i=1,\ldots,r$, of $\L_X$ such that the $e^i$ and $m_i$ span lagrangian sublattices with respect to $P$.  Write $P$ in this basis as the non-degenerate integral $2r\times 2r$ matrix
\begin{align}\label{JB}
    P &= \bpm & -B^t \\ B & \epm ,
\end{align}
where $B^i_{\ j} := P(e^i,m_j)$. 
In the dual basis $(m^i,e_i)$ of $\L_X^*$ (so $m^i(m_j)=\d^i_j$, $m^i(e_j)=0$, etc.) the special section is $s := a^D_i m^i + a^i e_i$, and $P(e_i,m^j) = -(B^{-1})^j_{\ i}$ so the induced pairing, $P^*= - P^{-1}$, is the inverse transpose of \eqref{JB} in the dual basis.
The $a_i$ are ``special coordinates" and the $a^D_i$ are ``dual special coordinates".
The metric is $ds^2 = 2 \Im ( da^D_i (B^{-1})^i_{\ j} d\bar a^j)$.
Flatness of $ds^2$ and the special coordinates, and the SK integrability of $s$ then imply that $a^D_j = \t_{jk} a^k$,
where $\t$ is a constant complex $r\times r$ matrix satisfying
$\t B = (\t B)^t$. 
Positivity of the metric implies $\Im(\t B)>0$.
These also imply that $a^D$ and $a$ are separately good holomorphic coordinates on the regular points of $\cC_\fg$.

In terms of the holomorphic family $X_u$ of abelian varieties over $\cC_\fg$ mentioned above, $\L_X$ and $P$ are its homology lattice and polarization, and $B\t$ is its complex modulus.

A monodromy $M \in \Sp_P(2r,\Z)$ of the EM charge lattice satisfies $MJM^t=J$, which implies that $M^{-t} P^* M^{-1} = P^*$, so $\Sp_{P^*}(2r,\Z) = \Sp_P(2r,\Z)$.
We define the dual EM monodromy by $M^* := M^{-t}$.
Upon continuing the special section along a closed path $\g$ in $\cC_\fg \setminus \{\text{singular locus}\}$, it can suffer a monodromy $M^*(\g) \in \Sp_{P^*}(2r,\Z)$.
This defines a monodromy map $\m^*: \pi_1(\cC_\fg \setminus \{\text{singular locus}\}) \to \Sp_{P^*}(2r,\Z)$, and we call $\im\m^* \subset \Sp_{P^*}(2r,\Z)$ the \emph{EM duality group} of the theory in question.
Constancy of $\t_{jk}$ then implies it is fixed by the EM duality group, so $m^*\t+n^* =  \t(p^*\t+q^*)$ for all $\bspm m^* & n^* \\ p^* & q^* \espm \in \im\m^*$.

In the $\cN{=}4$ case where the Coulomb branch is an orbifold by the Weyl group action, the EM duality group is the image of the Weyl group in $\Sp_{P^*}(2r,\Z)$ given by its action on the special section.
Furthermore, in this case $\t$ is only by EM duality invariant up to an overall complex constant which is the exactly marginal coupling constant of the $\cN{=}4$ sYM theory.

\subsection{Connection between special K\"ahler structures and Dirac pairing}\label{orbifoldSK}

To start with, let us choose as special coordinates $a^k = \a_k^*$ where $\a_k^*$ are is the dual basis of simple roots on $\C \otimes \ft^*$.  
(Dual simple roots are linearly independent real linear functions on $\ft^*$, so extend by linearity to good complex coordinates in neighborhoods of $\cC_\fg$ away from orbifold fixed points.)  
This corresponds to a choice of ``electric" lagrangian sublattice basis $\{e^k\}$ of $\L_X$ with $e^k = \a_k$, the simple roots.
Thus we are choosing the electric homology sublattice to be the root lattice.
Likewise we can choose as dual special coordinates $a^D_k = (\a^\v_k)^*$ (the dual basis of the simple co-roots), corresponding to the choice of ``magnetic" lagrangian sublattice basis the simple co-roots $\{m_k = \a^\v_k\}$, so the magnetic homology sublattice is the co-root lattice.
In this case it is clear that we have chosen the SK homology lattice to be
\begin{align}\label{vr-r}
    \L_X = \G^\v_r \oplus \G_r,
\end{align}
and so it coincides with the charge lattice \eqref{chrglatt} of the field theory.

A Weyl element $\s\in W_\fg$ acts on the electric charge lattice in this simple root basis by multiplication by an integral $r\times r$ matrix which we denote by the same symbol, $\s$.
Denote the matrix representation of the Weyl group element $\s\in W_\fg$ on this basis by $\s^\v$, as in \eqref{Weylact}.
Then $M_\s := \bspm \s^\v & 0\\ 0& \s \espm \in \GL(2r,\Z)$ for $\s\in W_\fg$ preserve the symplectic form $P$ \eqref{JB} with $B=-A_\fg$, the Cartan matrix of $\fg$.
This is just a restatement of the calculation of the last section around equations \eqref{Cartan1} and \eqref{Cartan2}.
Thus the monodromies $M_\s$ are in $\Sp_P(2r,\Z)$, so the above choices of special and dual special coordinates give a consistent SK structure on $\cC_\fg$ with polarization $P$, and this polarization coincides with Dirac pairing \eqref{Jfg} of the field theory.

But it is also clear that other, inequivalent, SK structures can be put on the $\cN{=}4$ Coulomb branch orbifold.  
The main consistency requirement is that the special section be chosen so that there is a basis in which its monodromies are given by integer matrices which preserve an integer symplectic form.  
Since the monodromies are induced by the action of the orbifolding Weyl group, $W_\fg$, such special sections will be related to sublattices of the weight plus co-weight lattice $\G_w^\v\oplus \G_w$ for $\fg$ which are preserved under $W_\fg$.  
As reviewed in appendix \ref{Cartmat}, for $\G_w$ these sublattices are the group lattices, which are in  1-to-1 correspondence with the different possible subgroups of the center of the simply connected gauge group; and similarly for $\G^\v_w$ as summarized in \eqref{g-lattices}.

So, for example, one can take special coordinates $a^k = \a_k^*$, the dual simple roots, as in the previous example, but choose the dual special coordinates differently to be $a^D_k = (\om^\v_k)^*$, where $\om_k$ are a basis of fundamental co-weights.
This corresponds to choosing the homology lattice
\begin{align}\label{vw-r}
    \L_X = \G^\v_w \oplus \G_r .
\end{align}
From the definition \eqref{Weylact} of the Weyl group actions it follows that if $\s_i$ is the matrix representation of the $\s_i$ action in the simple root basis, then $\s^\v_i$ acts in a fundamental co-weight basis as matrices $(\s_i)^{-t}$.  
Since the generating $\s_i\in W_\fg$ are reflections, $(\s_i)^{-t} = (\s_i)^t$, making their integrality apparent.
Thus the monodromy matrices $M_\s := \bspm \s^{-t} & 0\\ 0& \s \espm \in \GL(2r,\Z)$,and, furthermore, it follows immediately that they preserve a principal polarization
\begin{align}\label{Pvw-r}
    P = \bpm0 & 1 \\ -1  & 0 \epm.
\end{align}
This gives an example of an SK structure on $\cC_\fg$ in which the homology lattice is not the charge lattice.
Since the homology lattice is principally polarized, and since the charge lattice \eqref{vr-r} is a sublattice of \eqref{vw-r}, it is natural to guess that this SK structure corresponds to a choice of line lattice, and so to a choice of global structure of the field theory.

Indeed, it is easy to see that \eqref{vw-r} and \eqref{Pvw-r} correspond to the global structure $(\tG/Z(\tG))_0$ in the notation of \cite{Aharony:2013hda}, where the electric gauge group is the ``adjoint" group with trivial center, and the magnetic gauge  group is the simply-connected group $\til G^\v$.
This follows because $\G_r$ and $\G^\v_w$ are the group lattices of these two groups.
The zero subscript refers to the fact that the polarization \eqref{Pvw-r} on $\L_X$ is in block-skew form.
It is not hard to see that there are other SK structures in which,  for instance, the dual special coordinates are the fundamental co-weights shifted by multiples of weights according to their charges under the center $Z(\tG)$ of the gauge group.
This effectively shifts the generating monodromy matrices $M_\s$ to block triangular form which preserve a principal but non-block-skew polarization and which correspond to global forms $(\tG/Z(\tG))_n$ with $0\neq n\in Z(\tG)$.

SK structures corresponding to the other global forms can be constructed along similar lines.
They correspond to choices of Weyl-invariant sublattices of $\G^\v_w\oplus \G_w$ which preserve a principal integral symplectic pairing.
As was discussed in the last two sections and in appendix \ref{appC}, this is equivalent to the classification of global forms given in \cite{Aharony:2013hda}.

Note that it is also possible to construct SK structures ``intermediate" to the principally polarized line lattices and the ones with the physical Dirac pairing on the charge lattice.
The homology lattices which  occur in these cases could be interpreted as ``non-maximal" choices of mutually local line operators.
Their polarizations will be non-principal, and will have invariant factors which are divisors of those of the Dirac pairing.

Finally, $\cN{=}4$ theories have a 1-dimensional conformal manifold and an associated S-duality group.
As an abstract group, the S-duality group is the fundamental group of the conformal manifold in the orbifold sense.
Orbifold fixed points on the conformal manifold can be detected as  those values of the coupling where the effective theories on the Coulomb branch have an enhanced finite global symmetry group. 
If we take the coupling (a coordinate on the conformal manifold) to be $\t$ in the complex upper half plane, then the conformal manifolds for $\cN{=}4$ sYM theories are fundamental domains of the m\"obius action of a finite-index subgroup of $\SL(2,\Z)$ (or of a closely related group in the cases of $\fg=G_2$ or $F_4$ \cite{Argyres:2006qr}).

In general, the S-duality group cannot be unambiguously reconstructed from the SK geometry of the Coulomb branch.
The reason is simply that the Coulomb branch geometry only captures partial information about the field theory and so might present two Coulomb branch vacua with distinct physics as having isomorphic SK geometries.
So, if we try to reconstruct the conformal manifold by identifying values of $\t$ in the upper half plane with SK-isometric Coulomb branches, we will generally make mistaken identifications, leading to too small a conformal manifold.
Equivalently this will give too large an S-duality group, i.e., one which is a smaller-index subgroup of $\SL(2,\Z)$ than the physical S-duality group.

The choice of global structure (or choice of line lattice) of a field theory is an important datum distinguishing them.
As emphasized and explored in \cite{Aharony:2013hda}, the global structures of a sYM theory with given gauge algebra $\fg$ form orbits under S-duality transformations.
The S-duality group and conformal manifold thus depend on the global structure orbit.
For instance, if a given global structure formed a single orbit by itself, the S-duality group would be the full $\SL(2,\Z)$ and the conformal manifold would be a punctured sphere with a $\Z_2$ and a $\Z_3$ orbifold point.
(The puncture is the weak-coupling limit.)
An orbit involving more global structures will give a smaller S-duality group and a conformal manifold with a different set of orbifold points and punctures.

Since we have seen how to encode the choice of global structure in a choice of SK structure on the Coulomb branch, it follows that the S-duality groups and conformal manifold topologies that can be accessed from the Coulomb branch geometries with principally polarized homotopy lattices should correspond to the orbits found in \cite{Aharony:2013hda}.
The following subsections discuss from this point of view examples of Coulomb branch geometries for $\cN{=}2^*$ theories that have appeared in the literature.

\subsection{$\cN{=}2^*$ Coulomb branch geometries}\label{uNcurves}

We now discuss $\cN{=}2^*$ sYM Coulomb branch geometries from the perspective of understanding whether their homology lattices are charge lattices, line lattices, or something intermediate.  These geometries have been given in terms of Seiberg-Witten curves derived as spectral curves of integrable systems for $\fg=\su(N)$ in \cite{Donagi:1995cf} and for other simple $\fg$ in \cite{DHoker:1997hut, DHoker:1998xad}.

Naively, the Jacobian variety of a Riemann surface is principally polarized by its intersection pairing, and so one might think that the homology lattice derived from a Seiberg-Witten curve (family of Riemann surfaces) will be principally polarized.
By the dictionary worked out in the last subsection, this would seem to imply that the homology lattice of these SK geometries correspond to line lattices.
But this is not necessarily the case.
The Seiberg-Witten curves which appear in various constructions often have genus greater than the rank of the Coulomb branch.
As a result their homology lattices have too large a rank to be interpreted directly as either charge or line lattices.
Instead, an extra condition picking out an appropriate-rank sublattice of the homology lattice must be imposed.
The principal polarization of homology lattice, restricted to this sublattice, need no longer be principal.

For example, the integrable system \cite{Donagi:1995cf}, IIA/M theory brane \cite{Witten:1997sc}, and S-class $A_{N-1}$ \cite{Gaiotto:2009we, Gaiotto:2009hg} constructions of the $\cN{=}4$ $\su(N)$ sYM theory all describe a Seiberg-Witten curve, $\S_N$, which is a bouquet of $N$ tori all of complex modulus $\t$ and all identified at a marked point.  
This degenerate genus-$N$ curve is interpreted as the SW curve of the $\u(N)$ theory, whose Weyl group, $S_N$, acts by permuting the tori in $\S_N$. 
The rank of the Coulomb branch is $N{-}1$, so the homology lattice of $S_N$ has rank 2 greater than what is desired.

Indeed, the lattice $\L_X$ of homology 1-cycles on $\S_N$ is interpreted as the lattice of (possible) charges for the $\u(N)$ theory and its intersection form is the Dirac pairing. 
The basis $\{m_i, e^i,i=1,\ldots,N\}$ of the homology lattice, where $\{m_i,e^i\}$ is a canonical basis of the $i$th torus, has intersection form $P(m_i, e^j) = \d_i^j$.   

Now restrict to the sublattice of the homology lattice which corresponds to states neutral under the central $\u(1)$ factor of the $\u(N)$ gauge algebra, thereby identifying the effective homology lattice of the $\su(N)$ gauge theory SW curve.
We do this by identifying the rank-$2(N-1)$ symplectic sublattice of the rank-$2N$ homology lattice which is invariant under the Weyl group action.

An element $\pi\in S_N$ of the Weyl group acts as $\pi: \{m_i \, , \, e^i\} \mapsto \{m_{\pi(i)} \, , \, e^{\pi(i)}\}$. 
Decompose $\R^{2N} =  \R\otimes_\Z \L_X$ into invariant subspaces $\R \oplus \R \oplus \R^{N-1} \oplus \R^{N{-}1}$ under this action, where the first two factors are generated by $\sum_{i=1}^{N} m_i$ and $\sum_{i=1}^{N} e^i$, respectively, and the second two factors have bases $\{m_i {-} m_{i+1}\}$ and $\{e^i {-} e^{i+1}\}$ for $i=1, \ldots, N{-}1$, respectively. 
Though this decomposition is not unique, if we require that the decomposition is into symplectic subspaces with respect to the Dirac pairing, then we do get a unique decomposition $\R^{2N} \simeq \R^2 \oplus \R^{2(N-1)}$ where the $\R^{2(N-1)}$ factor is the sum of the last two factors of the previous decomposition.
Then the rank-$2(N{-}1)$ sublattice invariant under the Weyl group action is $\L_{\su(N)} := \L_X \cap \R^{2(N-1)}$ which has basis $\{ m_i {-} m_{i+1} \, , \, e^i {-} e^{i+1}\}$ for $i=1,2,\ldots,N{-}1$.
The pairing induced on the $\su(N)$ sublattice from the intersection pairing on $\L_X$, $P(m_i {-} m_{i+1} , e^j {-} e^{j+1} ) = 2\d_{i,j} - \d_{i+1,j} - \d_{i,j+1}$, is the Cartan matrix for $\su(N)$.

Thus the homology (sub)lattice and (induced) polarization of this SW curve are precisely those of the $\su(N)$ sYM charge lattice and Dirac pairing.
As a result, these SK structures do not encode any choice of global structure of the field theory.
We therefore expect that the S-duality group that is visible from this curve should be the full $\SL(2,\Z)$, and not one of the more refined subgroups associated to a given orbit of line lattices described in \cite{Aharony:2013hda}.
Indeed, the $\SL(2,\Z)$ S-duality of the curve is obvious from its initial description as a bouquet of identical tori all with one marked point and the same complex modulus.

The integrable system spectral curves for other simple Lie algebras $\fg$ \cite{DHoker:1998xad} have a similar structure. 
They give Riemann surfaces whose genus is the dimension of a non-trivial irreducible representation of the Lie algebra, so is always greater than the rank of the Lie algebra.
So it seems likely that their associated homology sublattices have non-principal induced polarizations.
It would be interesting to check whether these coincide with the charge lattices and Dirac pairings computed in the last section.

\subsection{Other curves for $\su(2)$ and $\su(3)$ $\cN{=}4$ sYM} \label{su2curves}

In low ranks the possible SK geometries are better understood.  
In rank 1 a full classification of these geometries is known \cite{Argyres:2015ffa, Argyres:2015gha, Argyres:2016xua}, and among them there are two corresponding to the $\su(2)$ $\cN{=}2^*$ theory.  
At rank 2 much less is known \cite{Martone:2021ixp}, but a second curve --- besides the one discussed in the previous subsection --- describing the $\su(3)$ $\cN{=}2^*$ theory is known.
These additional curves are precisely the ones expected to encode the global structure of the field theory.

\paragraph{su(2).}

One curve is the $N=2$ specialization of the $\su(N)$ curves described in the last subsection.
It was written directly in terms of a genus-1 Riemann surface in the original paper \cite{Seiberg:1994aj} of Seiberg and Witten.
As argued in detail in \cite{Argyres:2015ffa, Argyres:2015gha}, this curve has non-principal homology lattice polarization with invariant factor $2$.
(At rank 1 the invariant factor appears as an overall normalization of the polarization so is somewhat subtle to identify correctly.)
This coincides with the Dirac pairing on the charge lattice.
And, indeed, upon turning on the $\cN{=}2^*$ mass deformation, the conformal singularity on the Coulomb branch splits into three singularities corresponding to IR free $\u(1)$ gauge theories with massless hypermultiplets of (magnetic, electric) charges $(1,0)$, $(1,-1)$, and $(0,1)$.
These charges span the whole homology lattice of the curve, and so show that the homology lattice and charge lattice coincide.
This is also consistent with the S-duality group visible from this curve being the full $\SL(2,\Z)$ group.

In \cite{Argyres:2015gha} a second curve with Coulomb branch consistent with the $\su(2)$ $\cN{=}2^*$ theory was constructed.
The homology lattice of this curve has principal polarization, so we expect it should be identified with the $\su(2)$ with a choice of line lattice.
In fact, the three inequivalent choices of line lattices form a single orbit under S-duality, so there should only be a single such principally polarized SK geometry.
This interpretation is borne out by a closer comparison of the charge and homology lattices.
Upon turning on the $\cN{=}2^*$ mass deformation, the conformal singularity splits into three IR free singularities with massless hypermultiplet charges $(1,0)$, $(1, -2)$ and $(0, 2)$ which span the charge lattice.
This is an index-2 sublattice of the homology lattice, with induced polarization with invariant factor 2.
Furthermore, the S-duality group visible in this SK structure is the index-3 subgroup $\G^0(2) \subset \SL(2,\Z)$ generated by $T^2$ and $TS$ \cite{Argyres:2015gha}.
This is the S-duality group predicted by the line lattice analysis \cite{Aharony:2013hda}.

\paragraph{su(3).}

Similarly, our analysis shows that in addition to the $\su(3)$ curve with homology lattice equal to the charge lattice and $\SL(2,\Z)$ S-duality group, there should be a second Seiberg-Witten curve with principally polarized homology lattice corresponding to the single S-duality orbit of line lattices of this theory.
In fact, such a curve can be constructed \cite{AM22}.
Despite the simplicity of the orbifold analysis given in section \ref{orbifoldSK}, the description in terms of a Seiberg-Witten curve is quite complicated.
The Coulomb branch has complex coordinates $(u,v) \in\C^2$ with scaling dimensions $(2,3)$, respectively.
The curve is a family of genus-2 Riemann surfaces depending holomorphically on these vevs and on an exactly marginal (dimensionless) coupling parameter, $\t$, given in hyperelliptic form by
\begin{align}
  y^2&=-\frac{1}{576(u^3-v^2)}\Big(729(2+\tau)u^6+972(\tau-10)u^5 x^2+864 u\,v^2\, x(27v+2\tau x^3)\\\nonumber
  &+144 u^2\,v\,x^2(27(5+\tau) v+4(2+\tau) x^3)+108u^4(27(-6+\tau)v\, x+4(10+\tau)x^4)+\\\nonumber
  &16v^2(729v^2+108\tau v\,x^3+16 x^6)-16u^3(729 v^2-54(10+3\tau)v\, x^3-4(\tau-2)x^6)\Big).
\end{align}
The Seiberg-Witten 1-form is
\begin{align}
\l &= (u x+v) \frac{dx}y .
\end{align}
It can be checked that the periods of this 1-form define a special section satisfying the SK integrability condition, and give rise to metric non-analyticities along a single irreducible $u^3=v^2$ subvariety of the Coulomb branch, matching the geometry expected from the orbifold construction.
Since it is given as a family of genus-2 curves, its homology lattice is principally polarized.
However, the $\cN{=}2^*$ mass-deformed version of this curve is not known, so we cannot directly verify that the charge lattice is an index-3 sublattice of the homology lattice.

\acknowledgments 
We would like to thank Michele Del Zotto, I\~naki Garcia-Etxebarria, and Yuji Tachikawa for enlightening conversations and comments. 
The work of PCA is partially supported by DOE grant DE-SC0011784.
The work of MR is partially supported by a U.\ Cincinnati Joiner Fellowship.
The work of MM is supported by STFC grant ST/T000759/1.

\appendix

\section{Properties of simple Lie algebras and groups}
\label{Cartmat}

We set notation and recall some basic facts about weight and co-weight lattices and Cartan matrices of simple Lie algebras, $\fg$.

For each $\fg$ there is a simply-connected compact Lie group, $\tG$.  Other compact Lie groups $G$ with Lie algebra $\fg$, are given by quotients of $\tG$ by various subgroups of its center, $Z(\tG)$.  
Only those irreducible representations of $\fg$ which represent $Z(\tG)/Z(G)$ by the identity exponentiate to give representations of a given global form $G$.

A Cartan subalgebra $\ft\subset\fg$ is a maximal commuting subspace of $\fg$ and is always of dimension $r=\text{rank}(\fg)$.  In a given irrep $R$, the representation matrices of $h\in\ft$ can be simultaneously diagonalized giving vectors $\l\in \ft^*$ of simultaneous eigenvalues so that $\l(h)$ is an eigenvalue of $R(h)$.  The set $\{\l\}$ are the weights of $R$, and their integral span generates a lattice $\G_R\subset\ft^*$, the weight lattice of $R$.  Here $\ft^*$ is the real linear dual of $\ft$ (i.e., the space of linear maps from $\ft$ to $\R$) and $\G^*$ will denote the lattice integrally dual to $\G$ (i.e., $\G$ is the space of linear maps from $\G^*$ to $\Z$).

The roots, $\{ \a \}$, are the non-zero weights of the adjoint representation of $\fg$.  One can choose (not uniquely) a subset of $r=\text{rank}(\fg)$ simple roots, $\{ \a_i, \ i=1,\ldots,r \}$, which are a basis of $\G_r$ and which separate the roots into two disjoint sets: the positive roots, which are those roots which can be written as non-negative integer linear combinations of the simple roots; and the negative roots, which are the negatives of the positive roots.

The group lattice, $\G_G$, is defined to be the union of the weight lattices for all irreps $R$ of $G$,  $\G_G := \cup_R \G_R$, (though, in fact, the union of only a finite number of irreps suffices).
The smallest (coarsest) possible group lattice is the root lattice, $\G_r$, which is the weight lattice of the adjoint irrep of $\fg$.  It occurs as the group lattice of the group $\tG/Z(\tG)$ which has trivial center.  The largest (finest) possible lattice is the weight lattice of $\fg$, $\G_w$, and is the group lattice of $\tG$. 

From these definitions it follows that the group lattice is intermediate between the root and weight lattices of $\fg$ and determines the center of $G$ by
\begin{align}\label{g-lattices}
\begin{matrix}
   &&\G_r & \subset & \G_G & \subset & \G_w & \subset &\ft^* \\
   &&\updownarrow* &&\updownarrow*&&\updownarrow* &\\
\ft&\supset &\G^\v_w &\supset &\G^*_G &\supset &\G^\v_r &\\
\end{matrix}
\text{ with } \quad
Z(G) \ = (\G_G/\G_r)^* \ = \G^\v_w/\G^*_G ,
\end{align}
where the lattices connected by vertical arrows are integrally dual.  
$\G^\v_w$ and $\G^\v_r$ are the co-weight and co-root lattices, respectively, also called magnetic lattices.  The Goddard-Nuyts-Olive (GNO) or Langlands dual Lie algebra, $\fg^\v$, satisfies $\G_w(\fg) \simeq \G^\v_w(\fg^\v)$ and $\G_r(\fg) \simeq \G^\v_r(\fg^\v)$ where $\fg^\v = \fg$ for $\fg = \su(n), \so(2n), E_n, F_4, G_2$, but $\sp(2n)^\v = \so(2n+1)$ and $\so(2n+1)^\v = \sp(2n)$.  

In addition to its linear structure, $\ft$ comes with a positive definite real inner product inherited from the Killing form on $\fg$: $(e,f):={\rm tr}({\rm ad}(e){\rm ad}(f))$ for $e,f\in\fg$.  Upon restricting to $\ft$, one finds that $(\m,\n) = \sum_{\a\in{\rm roots}} \a(\m)\a(\n)$ for $\m,\n\in\ft$.  This inner product is defined up to a single overall normalization for simple $\fg$.  Choosing a normalization, the inner product can be used to select a canonical identification between $\ft$ and its dual $\ft^*$.  In particular, to each $\l \in \ft^*$, define $\l^* \in \ft$ by $(\l^*,\f) := \l(\f)\quad \forall \f\in\ft$.  Likewise, $\ft^*$ inherits an inner product from $\ft$ via $(\l,\m) :=(\l^*,\m^*) = \l(\m^*)$. 

Co-roots, $\a^\v$, are defined by
\begin{align}\label{corootmap}
\a^\v := \frac{2\a^*}{(\a,\a)} , \qquad \a \in \text{roots.}
\end{align}
When $\a$ and $\b$ are roots, then $\b(\a^\v)$ are integers for all simple Lie algebras.
\begin{align}\label{cartanmatrix}
A_{ij} := \a_i(\a_j^\v) , \qquad \a_i \in \text{simple roots,}
\end{align}
are the elements of the $r\times r$ integer Cartan matrix of the algebra.  The Cartan matrices are
\begin{align}
    A_{\su(r+1)} &= \bpm 
    \ph{-}2 & -1 &  & & &\\
    -1 & \ph{-}2 & -1 & & &\\
    &\ddots&\ddots&\ddots& &\\
    & & -1 & \ph{-}2 & -1 &\\
    & & & -1 & \ph{-}2 & -1 \\
    & & & & -1 &\ph{-}2
    \epm , & 
    A_{G_2} &= \bpm 
    \ph{-}2 & -3 \\
    -1 & \ph{-}2 
    \epm ,\\
    A_{\so(2r+1)} &= A_{\sp(2r)}^t = \bpm 
    \ph{-}2 & -1 &  & & &\\
    -1 & \ph{-}2 & -1 & & &\\
    &\ddots&\ddots&\ddots& &\\
    & & -1 & \ph{-}2 & -1 &\\
    & & & -1 & \ph{-}2 & -2 \\
    & & & & -1 &\ph{-}2
    \epm , &
    A_{F_4} &= \bpm 
    \ph{-}2 & -1 &    & \\
    -1 & \ph{-}2 & -2 & \\
    & -1 & \ph{-}2 & -1 \\
    & & -1 &\ph{-}2 
    \epm , \nn\\ 
    A_{\so(2r)} &= \bpm 
    \ph{-}2 & -1 & & & & & \\
    -1 & \ph{-}2 & -1 & & & &\\
    & \ddots & \ddots & \ddots & & &\\
    & & -1 & \ph{-}2 & -1 & & \\
    & & & -1 & \ph{-}2 & -1&-1\\
    & & & & -1 & \ph{-}2 & \\
    & & & & -1 & & \ph{-}2
    \epm ,&
    A_{E_r} &= \bpm 
    \ph{-}2 & -1 & & & & & \\
    -1 & \ph{-}2 & -1 & & & &\\
    & \ddots & \ddots & \ddots & & &\\
    & & -1 & \ph{-}2 & -1 & & -1\\
    & & & -1 & \ph{-}2 & -1 &\\
    & & & & -1 & \ph{-}2 & \\
    & & & -1 & & & \ph{-}2
    \epm ,\nn
\end{align}
and
\begin{align}
    \det A_{\su(r+1)} &= r+1 ,&
    \det A_{\so(2r+1)} &= \det A_{\sp(2r)} = 2 , & 
    \det A_{\so(2r)} &= 4 ,\nn\\
    \det A_{E_r} &= 9-r ,& 
    \det A_{F_4} &= 
    \det G_2 =1 .
\end{align}
In general $\det A = |\G_w/\G_r|$, the index of the root lattice as a sublattice of the weight lattice.
Also, $A_{\fg^\v} = (A_\fg)^t$.

The charge lattices in \eqref{g-lattices} can be computed as follows.  The root lattice, $\G_r$, is the integral span of the simple roots $\{\a_i\}$.  The co-root lattice, $\G^\v_r$, is spanned by the simple co-roots $\{\a^\v_i\}$.  The weight lattice, $\G_w$ is spanned by the fundamental weights $\{\om_i\}$ defined by $\om_i(\a^\v_j) = \d_{ij}$.  Finally the co-weight lattice, $\G^\v_w$, is spanned by the fundamental co-weights $\{\om^\v_i\}$ defined by $\a_i(\om_j^\v)=\d_{ij}$, or, equivalently, by $\om_i^\v = 2\om_i^*/(\a_i,\a_i)$.

The Weyl group, $W(\fg)$, is the group of orthogonal transformations of $\ft$ generated by reflections $\s_i$ for each simple root $\a_i$ which fix the hyperplane $\a_i(\f)=0$ in $\ft$ and act as 
\begin{align}\label{Weylact}
\s_i^\v(\f) &:= \f-\a_i(\f)\, \a_i^\v, \quad  \text{for}\quad \f\in \ft , \nn\\
\s_i(\m) &:= \m - \m(\a_i^\v)\, \a_i, \quad\text{for}\quad \m\in\ft^*. 
\end{align}
The action on $\ft^*$ is defined so $\s_i(\m)(\f) = \m(\s_i^\v(\f))$.  $W$ permutes the roots and acts transitively on them, and also acts transitively on the set of bases of simple roots.

\section{Invariant factors of a symplectic form} \label{invfacapp}

The structure theorem for finitely generated modules over a principal ideal domain --- see e.g., ch.\ IV, sec.\ 6 of \cite{Hungerford:2003} --- applied to a symplectic matrix $J$ over the integers implies that there is a basis in which $J$ takes the unique form \eqref{sympbas} whose diagonal entries are the invariant factors satisfying the divisibility condition \eqref{ddiv}.  For a direct proof in this case, see e.g., lemma on p.\ 305 of \cite{Griffiths:1978}.

The Dirac pairings we found for the $\cN{=}4$ sYM theories were already in the block-skew form $J = \bspm 0 & A \\ -A^t & 0 \espm$.  In this case finding the invariant factors can be done by putting the $r\times r$ matrix $A$ in Smith normal form, i.e., finding $P,Q \in \GL(r,\Z)$ such that $PAQ=D$ with $D={\rm diag}\{d_1,\ldots,d_r\}$ and $d_i|d_{i+1}$.  Then $J$ is put into canonical symplectic form \eqref{sympbas} by the change of basis
\begin{align}\label{A1}
    J \to \bpm P & 0 \\ 0 & Q^t \epm J
    \bpm P^t & 0 \\ 0 & Q \epm .
\end{align}
The Smith normal form algorithm is described in \cite{Wiki:2022snf}.

An algorithm for putting a general integral symplectic matrix (i.e., not necessarily in block-skew form) into canonical symplectic form is described in \cite{Speyer:2009}.  We reproduce it here for use in appendix \ref{appC}.
\begin{itemize}
\item[(0)]   Start with a basis of the lattice, $\{a_i, i=1, ..., 2r\}$, and reorder it, if necessary, so that $J(a_1,a_2) > 0$.  (Such an ordering always exists since $J$ is assumed non-degenerate.)  
\item[(1)] Set $d_r :=J(a_1, a_2)$.  If $d_r | J(a_1, a_k)$ for all $k>2$, then go to step (3).
\item[(2)] Find the smallest $k$ for which $d_r \nmid J(a_1, a_k)$, and set $q := [J(a_1, a_k) /d_r]$ (the integer part of the quotient).  Now replace $a_2 \to a_k-q a_2$, and $a_k \to a_2$.  Go to step (1) with this new basis.
\item[(3)] If $d_r | J(a_2, a_k)$ for all $k>2$, then go to step (4).  Find the smallest $k$ for which $d_r \nmid J(a_2, a_k)$, and set $q := [J(a_2, a_k)/d_r]$.  Replace $a_1 \to a_k-q a_1$, and $a_k \to a_1$.  Go to step (1) with this new basis.
\item[(4)] Set $\he_r := a_1$, $\hm_r := a_2$, and $b_k := a_k - \frac{1}{d_r} J(a_1, a_k) a_2 + \frac{1}{d_r} J(a_2, a_k) a_1$ for $k>2$.  Note that $J(\he_r,\hm_r) = d_r$ and $J(\he_r,b_k) = J(\hm_r,b_k) = 0$.  Define a reduced rank sublattice with basis $a_k := b_{k-2}$ for $k=1,\ldots,2(r{-}1)$.  Go to step (0) with this new rank-$(2r{-}2)$ lattice.
\end{itemize}
Since $d_r$ decreases after each step (2) and (3), eventually step (4) will be reached and the rank of the problem will be reduced. Thus, the algorithm eventually stops, outputting a basis $\{\he_1, \hm_1, ..., \he_r, \hm_r\}$ of $\L$ in which $J$ is skew-diagonal with $J(\he_i,\hm_j) = d_i \d_{ij}$ and $J(\he_i,\he_j) = J(\hm_i,\hm_j) = 0$. 

This basis does not guarantee that $d_i \mid d_{i+1}$, so does not directly determine the invariant factors of the Dirac pairing. Applying the Smith normal form algorithm to the pairing matrix as in \eqref{A1}, does give a Dirac pairing in which $d_i \mid d_{i+1}$.  In particular, say $d_i \nmid d_{i+1}$ and set
\begin{align}\label{A2}
    d'_i & := \gcd(d_i,d_{i+1}), &
    d'_{i+1} & := d_i d_{i+1} /\gcd(d_i,d_{i+1}).
\end{align}
Define a new basis by replacing
\begin{align}
    \bpm\he_i\\ \he_{i+1}\epm &\to 
    \bpm \a & \b \\ \g & \d \epm 
    \bpm\he_i\\ \he_{i+1}\epm, & 
    \bpm\hm_i\\ \hm_{i+1}\epm &\to 
    \bpm 1 & 1 \\ \b\g & \a\d \epm 
    \bpm\hm_i\\ \hm_{i+1}\epm ,
\end{align}
with $\a$, $\b$, $\g$, and $\d$ integers satisfying
\begin{align}
    \a d_i + \b d_{i+1} &= d_i', &
    \g & = - d_{i+1}/d'_i, &
    \d & = d_i/d'_i,
\end{align}
which exist by the definition of $d_i'$ and which ensure that the basis change matrices are invertible over the integers.  Then in the new basis $J$ is skew diagonal with new skew-eigenvalues $d_i'$ and $d'_{i+1}$ given by \eqref{A2}.  By successive application of the substitutions \eqref{A2} applied to pairs of skew eigenvalues, they are eventually brought to the invariant factor form in which $d_i \mid d_{i+1}$, at which point the \eqref{A2} substitution no longer changes the eigenvalues.

\section{Maximal symplectic sublattices of $\L^J$}
\label{appC}

Given a Dirac pairing in canonical form
\begin{align}\label{C0}
    J &= \bpm d_r \e & & & \\ & d_{r-1} \e & & \\ 
    & & \ddots & \\ & & & d_1 \e \epm, &
    \e &:= \bspm 0 & \ph-1 \\-1 & 0 \espm ,
\end{align}
with respect to a basis $\vev{e_1, \ldots, e_{2r}}$ of a rank-$2r$ charge lattice $\L$, the dual lattice $\L^J$ has basis $\vev{e^*_1, \ldots, e^*_{2r}}$ where $e^*_{2i-1} = d_i^{-1} e_{2i-1}$ and $e_{2i}^* = d_i^{-1} e_{2i}$ for $i = 1, \ldots, r$.  With respect to this basis the induced pairing is 
\begin{align}
    J^* &= \bpm d_r^{-1} \e & & & \\ & d_{r-1}^{-1} \e & & \\ 
    & & \ddots & \\ & & & d_1^{-1} \e \epm .
\end{align}
Note that we have re-ordered the basis relative to the one used in \eqref{sympbas} both by interlacing the electric and magnetic basis elements, and by reversing their overall ordering so that the invariant factors are ordered from largest to smallest.  This re-ordering is convenient for making the following argument.

\paragraph{Characterization of sublattices.}

We want to find the maximal sublattices $\cL \subset \L^J$ on which the induced Dirac pairing is integral.  Such a maximal sublattice is a full-rank sublattice of index $|\L^J/\cL| = \det D$.  An elementary result --- see Ch.\ I, Thm.\ I and Corr.\ 1 of \cite{Cassels:1971} --- states that distinct full-rank sublattices $\cL \subset \L^J$ of index $\det D$ are in 1-to-1 correspondence with bases $\vev{\he_1,\ldots,\he_{2r}}$ of the form 
\begin{align}\label{C1}
 \bpm \he_1 \\ \he_2 \\ \vdots \\ \he_{2r} \epm &= 
 V \bpm e^*_1 \\ e^*_2 \\ \vdots \\ e^*_{2r} \epm, &
 V &= \bpm v_{11} & & & \\
 v_{21} & v_{22} & & \\
 \vdots & & \ddots & \\
 v_{2r,1} & v_{2r,2} & \cdots & v_{2r,2r} \epm,
\end{align}
where the entries of the lower triangular matrix $V$ satisfy
\begin{align}\label{C2}
    0 &\le v_{ab} \in \Z, & 
    \prod_{a=1}^{2r} v_{aa} &= \det D, &
    &\text{and}&
    v_{ba} &< v_{aa} \quad \text{for}\ b>a,
\end{align}
and every distinct such $V$ corresponds to a distinct sublattice.  We refer to such $V$'s as sublattice basis matrices.

The induced Dirac pairing on $\cL$ in this basis is given by $\hat J = V J^* V^t$.  Split the pairing and sublattice basis matrices into blocks as
\begin{align}\label{C5}
    J^* &= \bpm J^*_{2s} & \\ & J^*_{2r-2s} \epm, &
    V &= \bpm V_{2s} & \\ W & V_{2r-2s} \epm,
\end{align}
where the subscripts denote the sizes of the square blocks.  Then
\begin{align}\label{C6}
\hat J = \bpm V_{2s}J^*_{2s}V_{2s}^t &
V_{2s} J^*_{2s} W^t \\ W J^*_{2s} V_{2s}^t & 
\ WJ^*_{2s} W^t {+} V_{2r{-}2s} J^*_{2r{-}2s} V_{2r{-}2s}^t
\ \epm .
\end{align}
A necessary condition for $\cL$ to be a line lattice is that its induced Dirac pairing, $\hat J$, is integral, and so in particular we must have $\Z \ni \det(V_{2s}J^*_{2s}V_{2s}^t) = (\det V_{2s} \cdot \Pf J^*_{2s})^2$, and so $\det V_{2s} \cdot \Pf J^*_{2s} \in \Z$.  Since $\det V_{2s} = \prod_{a=1}^{2s} v_{aa}$ (since it is lower triangular) and $\Pf J^*_{2s} = \prod_{i=r-s+1}^r d_i^{-1}$, we learn that a necessary condition for $V$ to describe a sublattice with integral induced pairing is that
\begin{align}\label{C7}
    \biggl( \prod_{i=r-s+1}^r d_i \biggr)  \mid  
    \biggl( \prod_{a=1}^{2s} v_{aa} \biggr)
\end{align}
for all $s=1,\ldots,r$.  

\paragraph{Reduction in rank when some invariant factors are 1.}

Now consider the case that $d_i=1$ for $i\le r-s$.  This, together with \eqref{C7} and the constraint from \eqref{C2} that $\prod_{a=1}^{2r} v_{aa} = \prod_{i=1}^r d_i$, implies that $v_{aa}=1$ for $a > 2s$.  The constraints \eqref{C2} then imply $v_{ab}=0$ for $b>a>2s$, i.e., that 
\begin{align}\label{C7.5}
    V_{2r-2s} = I_{2r-2s}
\end{align}
in \eqref{C5}, where $I_{2r{-}2s}$ is the $2r{-}2s\times 2r{-}2s$ identity matrix.  Furthermore, the $J^*_{2r-2s}$ symplectic block defined in \eqref{C5} is integral by virtue of the assumption that $d_i=1$ for $i\le r-s$.  The condition that $\hat J$ is integral together with its decomposition in \eqref{C6} then implies that both $W J^*_{2r{-}2s} W^t$ and $V_{2s} J^*_{2s} W^t$ are integral.  Define the vectors
\begin{align}\label{C8}
 \left(
 \begin{array}{c}
 \he_1 \\ \vdots \\ \he_{2s} \\ \hline
 \hw_{2s{+}1} \\ \vdots \\ \hw_{2r}
 \end{array} \right)
 & := 
 \left(
 \begin{array}{ccc|ccc}
 \ \ & &\ \ \ |\!\!\!&\ \ \ &\ \ \ \ &\ \ \ \\
 & V_{2s}\ \ & & & 0 & \\ 
 & & & & & \\ \hline
 & & & & & \\
 & W & & & 0 & \\ 
 & & & & &
 \end{array} \right)
 \left(
 \begin{array}{c}
 e^*_1 \\ \vdots \\ e^*_{2s} \\ \hline
 e^*_{2s{+}1} \\ \vdots \\ e^*_{2r}
 \end{array} \right).
\end{align}
The first $2s$ are $\cL$ basis vectors already defined in \eqref{C1}, and the remaining ones, the $\hw_k$, are vectors in $\L^J$.  The integrality of  $V_{2s} J^*_{2s} W^t$ and $W J^*_{2r{-}2s} W^t$ can then be interpreted as
\begin{align}\label{C9}
    J^*(\he_a, \hw_k ) &\in \Z & 
    &\text{and}&
    J^*(\hw_k, \hw_\ell ) &\in \Z .
\end{align}
Since $\cL$ is defined to be a maximal sublattice of $\L^J$ such that the induced Dirac pairing is integral, \eqref{C9} implies that the vectors $\hw_k$ are in $\cL$.  They must therefore be able to be written as integral linear combinations of the $\he_a$ basis vectors,
\begin{align}\label{C10}
    \hw_k &= \sum_{a=1}^{2s} w_{k,a} \he_a, &
    k &> 2s,
\end{align}
for some integers $w_{k,a}$.  From \eqref{C5} and \eqref{C8}, $\hw_k = \sum_{b=1}^{2s} v_{k,b} e^*_b$ and $\he_a = \sum_{b=1}^a v_{a,b} e^*_b$, so by \eqref{C10} 
\begin{align}\label{C11}
    \sum_{b=1}^{2s} v_{k,b} e^*_b &= 
    \sum_{a=1}^{2s} \sum_{b=1}^a w_{k,a} v_{a,b} e^*_b
    =
    \sum_{b=1}^{2s} \left(\sum_{a=b}^{2s} w_{k,a} v_{a,b}\right) e^*_b, &
    k &> 2s.
\end{align}
The $e^*_{2s}$ term implies 
\begin{align}\label{C12}
    v_{k,2s} &= w_{k,2s} v_{2s,2s},& k&>2s.
\end{align}
But the sublattice basis matrix condition \eqref{C2} implies $0\le v_{k,2s} < v_{2s,2s}$ so the only integer solution to \eqref{C12} is $w_{k,2s}=0$.  Using this in \eqref{C11} then gives a similar 1-term equation for the $e^*_{2s{-}1}$ term setting $w_{k,2s{-}1}=0$, and repeating this leads to 
\begin{align}\label{C14}
    W=0 .
\end{align}

Using \eqref{C7.5} and \eqref{C14} in \eqref{C5}, we have shown that the sublattice basis must be given by a $V$ of the form
\begin{align}
    V = \bpm V' & \\ & I_{2r{-}2s} \epm,
\end{align}
and $V'$ is a sublattice basis matrix of reduced size.  Thus the determination of the distinct maximal line lattices as sublattices of $\L^J$ can be reduced to listing the possible $2s \times 2s$ matrices of the form \eqref{C1} satisfying \eqref{C2}.

\paragraph{Application to counting sYM line lattices.}

We now apply this to the determination of the maximal line lattices for the $\cN{=}4$ sYM theories.  The invariant factors of their Dirac pairings, found above and listed in \eqref{N4invfac}, all have $s=1$, except for those with gauge Lie algebra $\fg = \so(4n)$ which have or $s=2$.  So we need only find sublattice matrices $V$ of size $2\times2$ or $4\times4$.

For the $s=1$ cases the invariant factors are $\{1,\ldots,1,d_r\}$.  Then we need to classify $2\times2$ $V'$'s satisfying the constraints in \eqref{C2} and which give an integral induced Dirac pairing.  Thus
\begin{align}\label{C16}
    V' &= \bpm v_{11} & 0 \\ v_{21} & v_{22} \epm, &
    &\text{with} &
    v_{11} v_{22} &= d_r &
    &\text{and} &
    0 &\le v_{21} < v_{11} ,
\end{align}
where the $v_{ab}$ are all non-negative integers.  The induced pairing on this block is
\begin{align}
    J' = \frac1{d_r} V'\e (V')^t
    = \frac{v_{11} \, v_{22}}{d_r} \, \e
    = \e ,
\end{align}
which is therefore integral for all $V'$ in \eqref{C16}.  One counts the possible such $V'$ as in  \cite{Gruber:1997, Zou:2006}, giving the number $N_{\rm line}$ of distinct line lattices:
\begin{align}\label{C18}
    N_{\rm line} &= \prod_k \frac{p_k^{n_k+1}-1}{p_k-1}, &
    &\text{where}& 
    d_r &= \prod_k p_k^{n_k} &
    &\text{is its prime decomposition.}
\end{align}
In particular, for low values of $d_r$ we have
\begin{align}\label{C19}
    \arraycolsep=3pt
    \begin{array}{c|cccccccccccccccc}
d_r          &1&2&3&4&5&6 &7&8 &9 &10 &11 &12 &13 &14 &15 
&\cdots\\ \hline
N_{\rm line} &1&3&4&7&6&12&8&15&13&18 &12 &28 &14 &24 &24
&\cdots
    \end{array} .
\end{align}

For the $s=2$ cases the invariant factors are $\{1,\ldots,1,2,2\}$.  Then we need only classify $4\times4$ $V$'s satisfying the constraints in \eqref{C2} and \eqref{C7}, i.e.,
\begin{align}
    V' &= \bpm v_{11} & & & \\ v_{21} & v_{22} & & \\
    v_{31} & v_{32} & v_{33} & \\
    v_{41} & v_{42} & v_{43} & v_{44} \epm,  \qquad
    0 \le v_{ba} < v_{aa} 
    \quad \forall b\neq a, \nn\\
    & \qquad \text{with} \quad
    v_{11} v_{22} v_{33} v_{44} = 4
    \quad \text{and}\quad
    2 \mid v_{11}v_{22} .
\end{align}
We split these into 7 cases according to the possible solutions for the diagonal elements:
\begin{align}
V'_I &= \bpm 2 &\ph{v_{43}} &\ph{v_{43}}&\ph{v_{43}}\\ 
    v_{21} & 2 & & \\
    v_{31} & v_{32} & 1 & \\
    v_{41} & v_{42} & 0 & 1 \epm, &
V'_{II} &= \bpm 2 &\ph{v_{43}} & &\ph{v_{43}}\\ 
    v_{21} & 1 & & \\
    v_{31} & 0 & 2 & \\
    v_{41} & 0 & v_{43} & 1 \epm, &
V'_{III} &= \bpm 2&\ph{v_{43}}&\ph{v_{43}}&\ph{v_{43}}\\
    v_{21} & 1 & & \\
    v_{31} & 0 & 1 & \\
    v_{41} & 0 & 0 & 2 \epm, \nn\\
V'_{IV} &= \bpm \ 1\ \ & & &\ph{v_{43}}\\ 
    0 & 2 & & \\
    0 & v_{32} & 2 & \\
    0 & v_{42} & v_{43} & 1 \epm, &
V'_V &= \bpm \ 1\ \ & &\ph{v_{43}} &\ph{v_{43}} \\ 
    0 & 2 & & \\
    0 & v_{32} & 1 & \\
    0 & v_{42} & 0 & 2 \epm, &
V'_{VI} &= \bpm 4 &\ph{v_{43}}&\ph{v_{43}}&\ph{v_{43}}\\ 
    v_{21} & 1 & & \\
    v_{31} & 0 & 1 & \\
    v_{41} & 0 & 0 & 1 \epm, \nn\\
V'_{VII} &= \bpm \ 1\ \ &\ph{v_{43}} &\ph{v_{43}}&\ph{v_{43}}\\ 
    0 & 4 & & \\
    0 & v_{32} & 1 & \\
    0 & v_{42} & 0 & 1 \epm,
\end{align}
where the undetermined $v_{ab} =0$ or $1$.  Now compute the induced sublattice pairings, $\hat J = V' J^* (V')^t$ and demand that all entries are integers.  This eliminates cases $VI$ and $VII$ and constrains the allowed values of the $v_{ab}$ in the other cases to
\begin{align}
V'_{Ia} &= \bpm 2 &\ph{v_{43}} &\ph{v_{43}}&\ph{v_{43}}\\ 
    0 & 2 & & \\
    v_{31} & 1 & 1 & \\
    1 & v_{42} & 0 & 1 \epm, &
    &\text{with} &
    v_{31} v_{42} &=0, &
    &\text{(3 lattices)} \\
V'_{Ib} &= \bpm 2 &\ph{v_{43}} &\ph{v_{43}}&\ph{v_{43}}\\ 
    0 & 2 & & \\
    0 & v_{32} & 1 & \\
    v_{41} & 0 & 0 & 1 \epm, &
    &\text{with} &
    v_{32} v_{41} &=0, &
    &\text{(3 lattices)} \nn\\
V'_{II} &= \bpm 2 &\ph{v_{43}} & &\ph{v_{43}}\\ 
    v_{21} & 1 & & \\
    0 & 0 & 2 & \\
    0 & 0 & v_{43} & 1 \epm, &
    &&&& &\text{(4 lattices)} \nn\\
V'_{III} &= \bpm 2&\ph{v_{43}}&\ph{v_{43}}&\ph{v_{43}}\\
    v_{21} & 1 & & \\
    0 & 0 & 1 & \\
    0 & 0 & 0 & 2 \epm, &
    &&&& &\text{(2 lattices)} \nn\\
V'_{IV} &= \bpm \ 1\ \ & & &\ph{v_{43}}\\ 
    0 & 2 & & \\
    0 & 0 & 2 & \\
    0 & 0 & v_{43} & 1 \epm, &
    &&&& &\text{(2 lattices)} \nn\\
V'_V &= \bpm \ 1\ \ & &\ph{v_{43}} &\ph{v_{43}} \\ 
    0 & 2 & & \\
    0 & 0 & 1 & \\
    0 & 0 & 0 & 2 \epm , &
    &&&& &\text{(1 lattice)} \nn
\end{align}
for a total of 15 maximal symplectic sublattices.  The induced pairings are
\begin{align}
J'_{Ia} &= \bpm
    0& 2& 1& v_{42} \\ 
    -2& 0&-v_{31}&-1 \\
    -1& v_{31}& 0& 0 \\
    -v_{42}& 1& 0& 0 \epm, &
J'_{Ib} &= \bpm
    0& 2& v_{32}& 1 \\ 
    -2& 0& -1& -v_{41} \\
    -v_{32}& 1& 0& 0 \\
    -1& v_{41}& 0& 0 \epm, &
  J'_{II\text{-}V} &= \bpm
    0& 1& 0& 0 \\ 
    -1& 0& 0& 0 \\
    0& 0& 0& 1 \\
    0& 0& -1& 0 \epm, \nn
\end{align}
which are manifestly principal for cases $II$-$V$.  They are also principal for cases $Ia$ and $Ib$ --- they must be by construction --- as can be checked by noting that their determinants are 1 or by computing the invariant factors as outlined in appendix \ref{invfacapp}.

\bibliographystyle{JHEP}

\end{document}